# PROTECTION AGAINST THE MAN-IN-THE-MIDDLE-ATTACK FOR THE KIRCHHOFF-LOOP-JOHNSON(-LIKE)-NOISE CIPHER AND EXPANSION BY VOLTAGE-BASED SECURITY [1]

L. B. KISH

Department of Electrical Engineering, Texas A&M University, College Station, TX 77843-3128, USA

ABSTRACT. It is shown that the original Kirchhoff-loop-Johnson(-like)-noise (KLJN) cipher is naturally protected against the man-in-the-middle (MITM) attack, if the eavesdropper is using resistors and noise voltage generators just like the sender and the receiver. The eavesdropper can extract zero bit of information before she is discovered. However, when the eavesdropper is using noise current generators, though the cipher is protected, the eavesdropper may still be able to extract one bit of information while she is discovered. For enhanced security, we expand the KLJN cipher with the comparison of the instantaneous voltages via the public channel. In this way, the sender and receiver has a full control over the security of measurable physical quantities in the Kirchhoff-loop. We show that when the sender and receiver compare not only their instantaneous current data but also their instantaneous voltage data then the zero-bit security holds even for the noise current generator case. We show that the original KLJN scheme is also zero-bit protected against that type of MITM attack when the eavesdropper uses voltage noise generators, only. In conclusion, within the idealized model scheme, the man-in-the-middle-attack does not provide any advantage compared to the regular attack considered earlier. The remaining possibility is the attack by a short, large current pulse, which described in the original paper as the only efficient type of regular attacks, and that yields the one bit security. In conclusion, the KLJN cipher is superior to known quantum communication schemes in every respect, including speed, robustness, maintenance need, price and its natural immunity against the man-in-the-middle attack.

## 1. Introduction: totally secure communication without quantum information with Kirchhoff-loop and Johnson(-like) noise

Recently, a totally secure classical communication scheme was introduced [1,2] utilizing two pairs of resistors and noise voltage generators, the physical properties of an idealized Kirchhoff-loop and the statistical physical properties thermal noise. In the idealized scheme of the Kirchhoff-loop-Johnson-(like)-noise (KLJN) cipher, the passively observing eavesdropper can extract zero bit of information. The *intrusive eavesdropper*, who emits a large and short current pulse in the channel, can extract *only one bit* of information while she is getting discovered [1] because the sender and the receiver are measuring their instantaneous current amplitudes and compare them via a public channel (like non-jamable radio transmission). The issue of the man-in-the-middle (MITM) attack was not studied in the original paper because of the author's earlier belief that no core physical secure layer is protected against such an attack and any security against the MITM attack has to be provided by additional (mainly software) tools, signatures, trusted third party, etc.

---

[1] This preprint was uploaded at arxiv.org on December 19, 2005; http://arxiv.org/abs/physics/0512177

In this paper, it is shown that the originally proposed KLJN scheme [1] is naturally protected against the MITM attack. Moreover, we propose an easy enhancement of the original security by the exchange and comparison of not only the instantaneous current amplitudes but also the instantaneous voltage amplitudes. This step further enhances the security against the MITM attack and it can be beneficial against other types of attacks of any practical realizations.

**2. General clarifications about the issue of security of physical secure layers**

In secure communication, any one of the following cases implies absolute security, however the relevant cases for physical secure layers are points 3 and 4:

1. The eavesdropper cannot physically access the information channel.

2. The sender and the receiver have a shared secret key for the communication.

3. *The eavesdropper has access and can execute measurements on the channel but the laws of physics do not allow extracting the communicated information from the measurement data.*

4. *The eavesdropper can extract the communicated information however, when that happens, it disturbs the channel so that the sender and receiver discover the eavesdropping activity.*

Keeping points 3 and 4 in mind, we can classify the focus topics of research of physical secure layers as follows:

*i) Absolute security of the idealized situation*. This is the most fundamental scientific part of the research and the mathematical model of the idealized physical system is developed and tested. The basic question is that how much information can be extracted from the data by the physical measurements allowed in the idealized situation? The original paper and the present study aim the investigation of this question.

*ii) Absolute security of the practical situation*. This part of the research requires an interdisciplinary effort including the fields of physics, engineering and data security. Because no real system can totally match the physical properties of the ideal mathematical model system, this kind of absolute security *does not exist in reality*; it is rather approached by an *only practically absolute* security. For example, in quantum communication, we have no ideal single photon source, no noise-free channel, and no noise-free detectors, and any of these deficiencies compromise absolute security. Similarly, with the KLJN cipher, cable resistance [3] and cable capacitance can cause information leak because the eavesdropper can execute measurements along the cable. This effect can be controlled and minimized by the particular design and choosing proper driving resistances and noise bandwidth that prohibit to make an acceptable statistics about the deviation of the noise strength along the cable within the clock period. Similarly, fast switching of the resistors can violate the *no-wave bandwidth rule* described in [1] (Eq. 9) but this problem can be avoided by using slow switches and/or filters at the line input.

*iii) If the code is broken, how many bits can be extracted by the eavesdropper before she is discovered due to the disturbance of the channel?* This question can also be treated at both the idealized-fundamental level and at the practical one. Some answers for the idealized case: RSA: infinite number of bits; Quantum: 20 - 10000 bits; KLJN cipher: 1 bit.

In the rest of the paper, we strictly focus on question *i)* while we are discussing the MITM attack in terms of point 4 above.

### 3. Security of the KLJN cipher against the man-in-the-middle attack and expanding the system with voltage based security

In this section we show that the original KLJN cipher arrangement [1] is secure against the MITM attack. The goal is to study the idealized model and show that the eavesdropper is discovered when she executes the attack. Finally, we propose a simple expansion of the scheme by comparing the voltage data, too, for enhanced security.

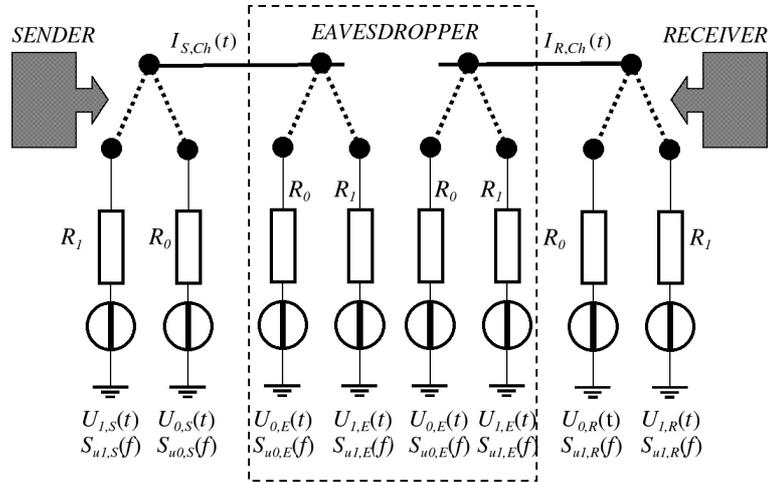

**Figure 1**. Man-in-the-middle-attack by using resistors with the same values and noise voltage generators with the same parameters as those of the sender and the receiver.

Figure 1 shows the MITM attack by using resistors with the same values and noise voltage generators with the same parameters as those of the sender and the receiver. The eavesdropper breaks the line at the middle and installs two KLJN communicators, one for the sender and another one for the receiver. According to the original scheme, the instantaneous current amplitudes, at the sender's end and at the receiver's end, are compared via a public channel. Because the eavesdropper's noise generators are different representations of the corresponding stochastic processes, their instantaneous amplitudes are different from those of the sender's and the receiver's noise generators. Therefore, the current amplitudes are different in the two loops and the eavesdropper is discovered within the reciprocal of the bandwidth, within the time resolution of the communicated noise, before extracting a single bit of information. We remind the reader [1] that the extraction of information from noise needs the making of a (short range) statistics through an averaging time window, which is the clock period, thus a single time (noise) sample provides zero information.

Mathematically, we can write as follows. Let $i \in 0,1$ ; $m \in 0,1$ ; $k \in 0,1$ ; $p \in 0,1$ . Then all the following voltages $U_{i,S}(t)$, $U_{m,E}(t)$, $U_{k,R}(t)$ and $U_{p,E}(t)$, see Figure 1, are statistically independent Gaussian stochastic processes with zero mean. The current at the sender's end and the receiver's end can be written as

$$I_{S,Ch}(t) = \frac{U_{i,S}(t) - U_{m,E}(t)}{R_i + R_m} \quad \text{and} \quad I_{R,Ch}(t) = \frac{U_{k,R}(t) - U_{p,E}(t)}{R_k + R_p}, \quad (1)$$

respectively. As the most pessimistic case, let us suppose that the denominators are equal and the nominators also contain the same type of noise generators, for example, $i = k$ and $m = p$. Then the RMS value of $I_{S,Ch}(t)$ and that of $I_{R,Ch}(t)$ are equal. Then the probability $P_0$ that $I_{S,Ch}(t_0) = I_{R,Ch}(t_0)$ at a given time moment $t_0$ is roughly equal to the ratio of the amplitude resolution $\Delta$ of the measurement system and the RMS value of these currents:

$$P_0 \approx \frac{\Delta}{I_{RMS}} \quad (2)$$

Let us suppose 7 bits resolution of the measurement (a pessimistic value), then $P_0 = 1/128$, which is less than 1% chance of staying hidden. On the other hand, $P_0$ is the probability that the eavesdropper can stay hidden during the correlation time $\tau$ of the noise, where $\tau$ is roughly the inverse of the noise bandwidth. Because the KLJN cipher works with statistics made on noise, the actual clock period $T$ is $N \gg 1$ times longer than the correlation time of the noise used [1]. Thus, during the clock period, the probability of staying hidden is:

$$P_{clock} = P_0^N \quad (3)$$

Supposing a practical $T = 10\tau$ (see [1]) the probability at the other example $P < 10^{-20}$.

This is the estimated probability that, in the given system the eavesdropper can extract a single bit without getting discovered. The probability that she can stay hidden while extracting 2 bits is $P < 10^{-40}$, for 3 bits it is $P < 10^{-60}$, etc. In conclusion, we can safely say that the eavesdropper is discovered immediately before she can extract a single bit of information. The probability of staying hidden can be estimated in the same manner in the rest of this paper, however we skip these calculations because of their trivial nature, and the straightforward considerations below.

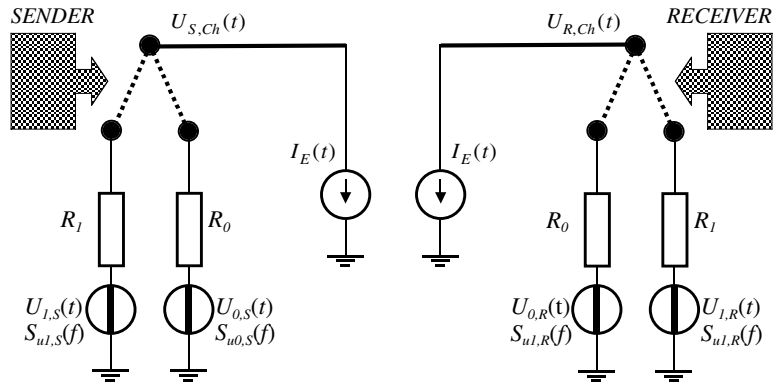

**Figure 2**. Man-in-the-middle-attack by using twin noise current generators with the same instantaneous current amplitudes.

Because the defense described above is based on the difference of the instantaneous current amplitudes at the two sides, a natural question arises. Can we provide security when the eavesdropper executes the MITM attack by using *two noise current generators* with the same instantaneous current amplitudes and RMS values imitating a reasonable noise current in the channel? Figure 2 shows the MITM attack by using twin noise current generators with the same instantaneous current amplitudes. In this case, the equality of the current amplitudes at the two sides is guaranteed by the twin noise current generators, therefore comparing the current amplitudes cannot be used for protection. For the best protection, we can expand the original KLJN cipher [1] by a voltage-based security enhancement. The sender and the receiver can compare the instantaneous voltage amplitudes at their end, via a public channel. Because the sender's noise generator and receiver's one have either different parameters or they are different representations of the same stochastic processes, their instantaneous amplitudes are different at most of the time. Therefore, the voltage amplitudes are different at the two ends and the eavesdropper is discovered practically immediately, before extracting a single bit of information.

However, it is important to note that the original scheme is also secure against this last kind of attack though not at the zero-bit security level but at the one-bit level. Normally the power density spectrum $S_{u,Ch}(f)$ of the voltage noise in the channel is *smaller* than that of the actual noise voltage generator $S_u(f)$ of the sender and the receiver, respectively, because it is proportional to the parallel resultant of those resistances:

$$S_{u,Ch}(f) < S_u(f) \tag{4}$$

However, during the MITM attack shown in Figure 2, the following relations hold:

$$S_{u,Ch}(f) = S_u(f) + R^2 S_{i,E}(f) \quad \text{therefore} \quad S_{u,Ch}(f) > S_u(f) \, . \tag{5}$$

That means, the MITM attack will be discovered after the sufficient statistics is made on the spectrum during the clock period. During the communication of the secure bit, the side using the larger resistor will discover the attack faster. At the same time, the eavesdropper, while getting discovered, can extract one bit of information. Thus, the expansion by the voltage-based security measure described above provides an enhanced security.

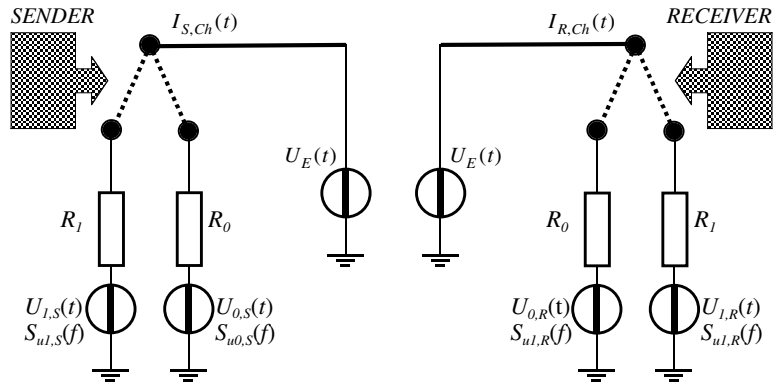

**Figure 3.** Man-in-the-middle-attack by using twin noise voltage generators with the same instantaneous current amplitudes.

Because the defense described in the context of Figure 2 is based on the difference of the voltages at the two sides, a natural question arises. Can we provide security when the eavesdropper executes the MITM attack by using *two noise voltage generators* with the same instantaneous voltage amplitudes and RMS values imitating a reasonable noise voltage in the channel? Figure 3 shows the man-in-the-middle-attack by using twin noise voltage generators with the same instantaneous current amplitudes. In this case, the twin noise voltage generators guarantee the equality of the instantaneous voltage amplitudes at the two sides. Therefore comparing the voltage amplitudes cannot be used for protection. For the protection, the sender and the receiver can compare the instantaneous current amplitudes, via a public channel. Because the sender's noise generator and receiver's one have either different parameters or they are different representations of the same stochastic processes, their instantaneous amplitudes are different. Therefore the current amplitudes are different in the two loops and the eavesdropper is discovered immediately, before extracting a single bit of information.

## 4. Conclusion

The KLJN cipher with public channel for comparing currents [1] is naturally protected against the MITM attack and the eavesdropper is discovered with a very high probability while or before she can extract a single bit. Enhanced security can be reached by comparing the voltages and then the eavesdropper is discovered with a very high probability before she can extract a single bit of information. Thus within the idealized model scheme, the man-in-the-middle-attack does not provide any advantage compared to the regular attack considered earlier. The remaining possibility is the attack by a short, large current pulse, which described in the original paper [1] as the only efficient type of regular attacks, and that yields the one bit security.

Therefore, the KLJN cipher is superior to known quantum communication schemes in every respect, including speed, robustness, maintenance need, price and its natural immunity against the man-in-the-middle attack.

**Acknowledgments**

Pre-published on the arxiv.org server on December 19, 2005 as http://arxiv.org/abs/physics/0512177 . Discussions with a number of security experts and scientists/engineers challenging the security of the KLJN scheme and emphasizing the importance of the MITM attack are appreciated. Without the ability to provide a complete list, here are some of the most significant commenters: Jonathan Wolfe, Bruce Schneier, Steven M. Bellovin, David Wagner, Charlie Strauss, Adrian Cho, Janos Bergou, Julio Gea-Banacloche, Matthew Skala, Greg Kochanski, Christian Vandenbroeck, Frank Moss, Derek Abbott, Bob Biard.